\documentstyle[12pt]{article}

\newlength{\dinwidth}
\newlength{\dinmargin}
\begin{document}

\def\bold#1{\setbox0=\hbox{$#1$}%
     \kern-.025em\copy0\kern-\wd0
     \kern.05em\copy0\kern-\wd0
     \kern-.025em\raise.0433em\box0 }
\def\slash#1{\setbox0=\hbox{$#1$}#1\hskip-\wd0\dimen0=5pt\advance
       \dimen0 by-\ht0\advance\dimen0 by\dp0\lower0.5\dimen0\hbox
         to\wd0{\hss\sl/\/\hss}}
\def\lq{\left [}
\def\rq{\right ]}
\def\LL{{\cal L}}
\def\VV{{\cal V}}
\def\AA{{\cal A}}

\newcommand{\be}{\begin{equation}}
\newcommand{\ee}{\end{equation}}
\newcommand{\bea}{\begin{eqnarray}}
\newcommand{\eea}{\end{eqnarray}}
\newcommand{\nn}{\nonumber}
\newcommand{\dd}{\displaystyle}
\newcommand{\bra}[1]{\left\langle #1 \right|}
\newcommand{\ket}[1]{\left| #1 \right\rangle}
\thispagestyle{empty}
\vspace*{1cm}
\rightline{BARI-TH/93-139}
\rightline{UTS-DFT-93-14}
\rightline{April 1993}
\vspace*{2cm}
\begin{center}
  \begin{Large}
  \begin{bf}
          Phenomenology of  $B_s$ decays\\
  \end{bf}
  \end{Large}
  \vspace{8mm}
  \begin{large}
P. Blasi $^{a}$, P. Colangelo $^{a,}$ \footnote{e-mail address:
COLANGELO@BARI.INFN.IT},  G. Nardulli $^{a,b}$\\
  \end{large}
  \vspace{6mm}
$^{a}$ Istituto Nazionale di Fisica Nucleare, Sezione di Bari, Italy\\
  \vspace{2mm}
$^{b}$ Dipartimento di Fisica, Universit\'a
di Bari, Italy \\

\vspace*{8mm}
  \begin{large}
N. Paver\\
  \end{large}
  \vspace{6mm}
Dipartimento di Fisica Teorica, Universit\'a di Trieste, Italy \\
  \vspace{2mm}
Istituto Nazionale di Fisica Nucleare, Sezione di Trieste, Italy\\
\end{center}
\begin{quotation}
\vspace*{1.5cm}
\begin{center}
  \begin{bf}
  ABSTRACT
  \end{bf}
\end{center}
\vspace*{0.5cm}
Using the QCD sum rules technique
we study several aspects of the phenomenology of the b-flavoured strange meson
$\dd \overline  {B_s^0}$. In particular, we evaluate the mass of the particle,
 the leptonic constant and the form factors of the decays
$\dd \overline {B_s^0} \to D^{+}_s \ell^- \bar\nu$,
$\dd \overline  {B_s^0} \to D^{*+}_s \ell^- \bar\nu$,
$\dd \overline  {B_s^0} \to K^{*+} \ell^- \bar\nu$. We also calculate,
 in the factorization approximation, a number of  two-body non leptonic
$\dd \overline {B_s^0}$ decays.
Finally, we compare our evaluation of the $\dd SU(3)_F$ breaking
effects in the $\dd \overline {B_s^0}$ channel to other
estimates.\\
\noindent
\end{quotation}
\vspace*{1.cm} PACS n:13.20.Jf,13.25.+m

\newpage
\baselineskip=18pt
\setcounter{page}{1}
\section{Introduction}
The interest for the $b$-flavoured strange
meson ${\bar B_s^0}$ ($b{\bar s}$) has been recently prompted by the reported
evidence for the production of this particle in the hadronic $Z^0$ decays at
LEP \cite{aleph,delphi,opal}.
A signal of correlated $D_s^+ \ell^-$ pairs ($\ell=\mu,e$) has been observed
 \cite{foot1}, with the lepton having a large momentum and a large momentum
component with respect to the b quark direction;  this signal can be
attributed to the semileptonic process
\be {\bar B_s^0} \to D_s^+ \ell^- {\bar \nu_\ell} X \label{eq:decay}\ee
\noindent
which,  by analogy with the $B_{u,d}$ case,  is expected to occur at the $10\%$
level. The indication of $B_s$ mesons is confirmed by the
observation of an excess of inclusive $D_s^+$ production, whose measured value
is larger than the expected production from $B_{u,d}$.\par
Evidence for the $B_s$ production at $\Upsilon (5S)$ was already reported by
the CUSB Collaboration at CESR \cite{cusb}. Moreover, indications for $B_s$
have been deduced from the measurement of the rate of same sign dileptons at
 the
hadron $p{\bar p}$ colliders \cite{dileptons_pp}
and at LEP \cite{dileptons_lep}: since this rate is
larger than the corresponding quantity measured at $\Upsilon (4S)$
\cite{cleo1}, the
difference can be attributed to the presence of ${\bar B_s^0}$, $B_s^0$ mesons
with a (nearly) maximal mixing.\par
Ongoing measurements will soon provide us with a value for the mass difference
$m_{B_s}-m_{B_d}$ by reconstructing non leptonic decay channels;
as for the lifetime $\tau_{B_s}$, the measured value
\cite{delphi1}
\be \tau_{B_s}\;=\;1.1\pm 0.5\;ps \label{lifetime}\ee
\noindent
is still dominated by the statistical error, so that no information on the
possible role of non spectator effects in this channel is available yet.

{}From the  theoretical standpoint, the interest for the ${\bar B_s^0}$ meson
stems from the possibility of clarifying the size of
the light flavour $SU(3)_F$ breaking effects
in the $b$ quark sector. In the charm sector some hints on such effects can be
obtained by comparing $D^+$ and $D_s$; the difference \cite{pdg}
\be m_{D_s}-m_{D^+}\;=\;(99.5\pm 0.6)\;MeV \label{eq:diff}\ee
\noindent
shows that these effects are of the order of $5\%$ for the mass of the
particles. In the $b$ system the $SU(3)_F$ breaking terms, which account for
the deviations from unity of the ratios ${m_{B_s}/ m_{B_d}}$,
${f_{B_s}/ f_{B_d}}$, etc., play a significant role in the possibility of
constraining the Cabibbo-Kobayashi-Maskawa matrix and, consequently, the quark
sector of the Standard Model. As a matter of fact, within the Standard Model
the mixing between ${\bar B_s^0}$ and $B_s^0$ occurs with the parameter
$x_s=(\Delta M / \Gamma)_{B_s}$ given by \cite{mixing}
\be x_s={G_F^2\over 6 \pi^2} \tau_{B_s} m_W^2 m_{B_s} (f_{B_s}^2 B_{B_s})
\eta_{B_s} |V_{ts}^* V_{tb}|^2 y_t f_2(y_t) \hskip 5pt.\label{eq:parxs}\ee
\noindent
Eq. (\ref{eq:parxs}) shows that the ratio $x_s/ x_d$ is independent of
the (still unknown) top quark mass $m_t$; the experimental determination of
this ratio implies a measurement of $|V_{ts}/V_{td}|$ once $(f_{B_s}^2
B_{B_s})/ (f_{B_d}^2 B_{B_d})$, $m_{B_s}/  m_{B_d}$ and
${\tau_{B_s}/ \tau_{B_d}}$ have been calculated and (or) measured
\cite{ali} .

The ratios ${m_{B_s}/ m_{B_d}}$ and ${f_{B_s}/ f_{B_d}}$ are
available presently
from potential models for the quark-antiquark systems \cite{colangelo1}.
The quantity ${(f_{B_s}^2 B_{B_s})/ (f_{B_d}^2 B_{B_d})}$ has been
estimated also by using the Heavy Quark Effective Chiral Perturbation Theory
\cite{grinstein}.

In this paper we calculate
${m_{B_s}}$, ${f_{B_s}}$ and the ratios
${m_{B_s}/ m_{B_d}}$,
${f_{B_s}/ f_{B_d}}$ by QCD sum rules \cite{svz}. This method is
deeply rooted in the QCD framework of the strong interactions, and has been
successfully applied to different aspects of the light
\cite{shiflibro,reinders} and
heavy hadrons \cite{colangelo2}. It avoids the notion of
wave-function for a system of constituent
quarks,  and directly relates hadronic properties (masses,
leptonic constants, etc.) to fundamental QCD quantities like current quark
masses, $\alpha_s$, and a set of parameters, the "condensates", which describe
the deviations from the asymptotically free behaviour at short distances by
allowing the inclusion of a series of power corrections.

In the QCD sum rules
approach the $SU(3)_F$ breaking effects in the static  parameters of the
heavy mesons
can be systematically taken into
account. Moreover, this technique permits the calculation of
 a number of dynamical
heavy system properties, e.g. the form factors that describe the semileptonic
decays ${\bar B_s^0}\to D_s^+(D_s^{*+}) \ell^- \nu$
and their deviations from the analogous quantities related to
${\bar B_d^0}\to D^+(D^{*+}) \ell^- \nu$.

The plan of the paper is as follows. In section 2 we evaluate the mass
and the leptonic constant of the ${\bar B_s^0}$ meson by two-point function QCD
sum rules. An analysis of the ratios ${m_{B_s}/ m_{B_d}}$ and
${f_{B_s}/ f_{B_d}}$ allows us to estimate the size of
$SU(3)_F$ breaking in these quantities.
Since the calculation can be
extended in a straightforward way to the $D_s$ meson, we calculate
$f_{D_s}$ and compare our findings with a number of recent experimental
and theoretical determinations.
 By using three point function QCD sum rules
we calculate in section 3 the
hadronic matrix elements that describe the semileptonic
decays ${\bar B_s^0}\to D_s (D_s^*) \ell^- \nu$ and ${\bar B_s^0}\to
K^* \ell^- \nu$.
Also in this case we evaluate the light flavour symmetry breaking effects.
In section 4 we estimate, in the factorization
hypothesis, the width of several two-body non leptonic $B_s$ decays.

\vglue 2cm
\section{ $B_s$ mass and leptonic constant}
A number of estimates of the leptonic constants for the heavy-light quark
mesonic systems can be found in the literature.  In particular,
QCD sum rules have been used to evaluate
 the B meson leptonic constant $f_B$ both for a
finite \cite{reinders1,dominguez} and an infinite heavy quark mass $m_b$
\cite{shuryak,neubert}.
Here we apply this method to the calculation of $f_{B_s}$
defined by the matrix element
\be \bra{0} {\bar b} i \gamma_5 s \ket{{\bar B_s^0}}\;=\;{f_{B_s} m_{B_s}^2
\over m_b + m_s} \label{eq:lepconst}\ee
\noindent
($m_b$ and $m_s$ are the $b$ and $s$ quark masses).
As usual in the QCD sum rules approach, the starting point is the correlator of
quark currents:
\be \Pi (q^2)\;=\;i\int dx \; e^{iqx} \bra{0} T(J_5 (x) J_5^{\dag} (0)) \ket{0}
\label{eq:twopoint}\ee

\noindent
with $J_5={\bar b} i \gamma_5 s$.
This correlator can be evaluated in two
different ways. First, by a short-distance operator product
expansion in QCD ($q^2 \to - \infty$), which gives the
 perturbative (P) contribution, written through a dispersion relation
\be \Pi (q^2)\;=\;{1\over \pi} \int ds \; {\rho_{P} (s) \over s - q^2}
\label{eq:pertdisp1} \hskip 5pt , \ee
\noindent
and non perturbative (NP) power corrections parameterized by vacuum
matrix elements of
quark and gluon field operators. These terms are ordered according to
the dimension; they represent the breaking of asymptotic freedom.
Therefore, the QCD form of the correlator reads:
\bea
\Pi_{QCD} (q^2)\;&=&\;\Pi^{P} (q^2) + \Pi^{NP} (q^2)\;=\nn \\
&=&\;\Pi^{P} (q^2) + C_3 (q^2) <{\bar s}s> + C_4 (q^2) <{\alpha_s \over \pi}
G^2> + \nn \\
&+&\;C_5 (q^2) <{\bar s} g \sigma G s> + ...\label{eq:wilson}
\eea
The perturbative spectral function $\rho_P(s)$ is given to the lowest order in
$\alpha_s$ by
\be \rho_P(s)\;=\;{3\over 8 \pi} {\sqrt{\lambda(s,m_b^2,m_s^2)}
\over s}\lbrack s - (m_b - m_s)^2
\rbrack \; \Theta \lbrack s - (m_b + m_s)^2 \rbrack  \label{spectral}\ee
\noindent
where $\lambda$ is the triangular function;
the $O(\alpha_s)$ corrections can be found in Ref.\cite{reinders}. The
coefficients $C_3$,
$C_4$ and $C_5$ in eq.(\ref{eq:wilson}) can be calculated
using the fixed point
technique \cite{novikov} with the result:
\bea
C_3 &=&  {m_b\over q^2 - m_b^2}-{m_s\over 2}{q^2-2m_b^2\over
(q^2 - m_b^2)^2} + {m_s^2 m_b^3\over (q^2 - m_b^2)^3} \label{eq:d3}\\
C_4 &=&{1\over 12} {1\over{(q^2-m_b^2)}}
\Bigl[-1-6{{ m_s m_b q^2} \over{(q^2-m_b^2)^2}}\ln{{q^2-m_b^2\over m_s m_b}}
 \; + \nn \\
&+&{m_s \over m_b}\left(1+{8m_b^2\over{(q^2-m_b^2)}}+
{6m_b^4\over{(q^2-m_b^2)^2}}\right)\Bigr] \label{eq:d4} \\
C_5 &=& -{1\over 2}\big\lbrack  {m_b\over (q^2 - m_b^2)^2} +
{m_b^3\over (q^2 - m_b^2)^3}\big\rbrack \hskip 5pt . \label{eq:d5}
\eea
\noindent

Actually, the main contribution comes from the $D=3$ and $D=5$ terms.\par

The second evaluation of the correlator is obtained
by writing the
spectral function $\rho(s)$
in terms of hadronic (H) resonances and of a continuum of states;
assuming the dominance of the lowest lying
resonance, one writes:
\be \rho_H(s)\;=\;\pi ({f_{B_s} m_{B_s}^2 \over m_b + m_s})^2 \delta (s -
m_{B_s}^2) + \rho_{CONT} (s) \; \Theta (s - s_0) \hskip 5pt ,
\label{eq:resonance}\ee
\noindent
where $s_0$ is an effective  threshold which separates the
contribution of the resonance from the continuum.
According to duality,
the continuum spectral function can be modeled
as in perturbative QCD:
therefore in (\ref{eq:resonance}) $\rho_{CONT}
(s) = \rho_{P} (s)$.

In the QCD Sum Rules approach, a region in $q^2$ (duality window)
has to be found where the hadronic and the QCD expressions for the correlator
 match with each other.
The matching can be improved by  a Borel transformation
defined by the operator
\be
{\cal B} = {(-Q^2)^n \over (n-1)!} \Big( {d \over d Q^2} \Big)^n
\label{eq:borel}
\ee
\noindent in the limit
$Q^2 \to \infty$ ($Q^2=-q^2$),
$n \to \infty$ and ${Q^2 / n}=M^2$ fixed,
applied to
both the hadronic and QCD sides of the rule.
 One obtains:
\be {1\over \pi} \int ds \; \rho (s) \; {e^{-s/M^2}\over M^2} \;=\;
\Pi^{P}(M^2)
  + \Pi^{NP} (M^2) \hskip 5pt  \label{eq:bortrans}\ee
\noindent
and  a daughter sum rule
for the mass  of the meson by
differentiating eq.(\ref{eq:bortrans}) with respect to ${1/ M^2}$.

Let us discuss the values of the parameters appearing in the sum rule
eq.(\ref{eq:bortrans}). The strange quark mass $m_s$ and the strange quark
condensate $<{\bar s}s>$ are responsible for the deviation
of eq.(\ref{eq:bortrans})
from the
analogous expression for the $B$ meson. Both these parameters are fixed by the
analysis of the baryonic states given in ref.\cite{reinders2}:
$m_s=0.14 - 0.15\;GeV$ and
$<{\bar s}s>=0.8 <{\bar d}d>$ with
$<{\bar d}d>=(-0.23 \;GeV)^3$; the mixed $D=5$  condensate can be expressed
in terms of $<{\bar s}s>$:
$<{\bar s}g \sigma G s>= m_0^2 <{\bar s}s>$ with $m_0^2=0.8\; GeV^2$.

The (pole) mass of the $b$ quark plays a crucial role in the sum rule. We use
the value fixed in ref.\cite{reinders1} by analyzing the
$\Upsilon$ system (see also \cite{paver}): $m_b=4.6 - 4.7 \;GeV$ \cite{smilga}.

The last QCD input parameter is $\alpha_s$;
we use the value obtained at the scale
$m_b$ with $\Lambda_{QCD} = 150-200 \; MeV$.

There are now two quantities that must be fixed: the  effective threshold
$s_0$ and the duality
window in the Borel parameter $M^2$.
The range of acceptable $M^2$ values can be fixed by requiring a hierarchical
structure in the contributions of the OPE and in the resonance-continuum
hadronic side. On the other hand, the value of $s_0$ can be changed in a
small interval: we use $s_0=33\div 36\;GeV^2$.
The typical curves are depicted in fig.1, where the duality region is
also shown.
Our result is:
\bea
m_{B_s} &=& (5.4\pm 0.1)\;GeV\nn \\
f_{B_s} &=& (190\pm 20)\;MeV  \label{eq:fbs}
\eea
\noindent where the uncertainties are due to the variation of the
parameters in their allowed intervals.

Before discussing these results let us observe that the same
calculation can be performed for the $D_s^+$ ($c \bar s$) meson. Using
$m_c=1.35 \; GeV$, $s_0=6-7 \; GeV^2$  and $\alpha_s$ at
the scale $m_c$ we get:
\bea
m_{D_s} &=& (2.0\pm 0.1)\;GeV\nn \\
f_{D_s} &=& (195\pm 20)\;MeV \hskip 5pt. \label{eq:fds}
\eea

\noindent
Within the uncertainties
the  result for the leptonic constant  is compatible
with the value obtained in
ref.\cite{abada} by a numerical calculation on the lattice:
$f_{D_s} = (230 \pm 50)\;MeV \hskip 5pt.$ Moreover, it is in agreement with
the measurement of the WA75 Collaboration \cite{wa75}
\be f_{D_s} = (232 \pm 45 \pm 20 \pm 48)\;MeV \hskip 5pt \label{eq:wa75} \ee
\noindent
obtained by the observation of leptonic decays $D_s^+ \to \mu^+ \nu$
in emulsion. Another estimate of
$f_{D_s}$ has been given in \cite{argus2,cleo2}
using the non leptonic decay channel $B \to D (D^*) \; D_s^+$ and the
factorization hypothesis, with a similar result.

As stated above, the uncertainties in eqs.(\ref{eq:fbs},\ref{eq:fds}) are due
to the
variation of $s_0$ and $M^2$ in the stability window.
Trying to reduce this error (mainly in the prediction of $m_{B_s}$) we have
studied the ratios ${m_{B_s}\over m_{B_d}}$ and
${f_{B_s}\over f_{B_d}}$ by writing the ratios of the corresponding
rules with two different continuum thresholds $s_0$ ($33\div36$ $GeV^2$ for
$B_s$ and $32\div 35$ $GeV^2$ for $B$).
These quantities display a softer dependence on the
parameters and are remarkably stable in $M^2$ as shown in fig.2.
This allows us to predict:
\be {m_{B_s}\over m_{B_d}}\;=\;1.005\;\pm\;0.002\label{eq:mratio}\ee
\be {f_{B_s}\over f_{B_d}}\;=\;1.09\;\pm\;0.03\label{eq:fratio}\ee
\noindent
with the uncertainty reduced by a factor of 2 with respect to
eqs.(\ref{eq:fbs}). The conclusion is
that the size of $SU(3)_F$ breaking effects
are of $0.5\%$ for the $B_s$ mass and less than $10\%$  for the
leptonic constant; these effects mainly come from the value of the
$<\bar s s >$ condensate.

\section{Semileptonic form factors}
The hadronic matrix elements of the transitions ${\bar B_s^0}\to P^+ e^- {\bar
\nu_e}$ and ${\bar B_s^0}\to V^+ e^- {\bar \nu_e}$ ($P$ and $V$ are strange
pseudoscalar and vector mesons, respectively) can be written in terms of form
factors using the decomposition in Ref. \cite{WSB}:
\bea
\bra{P^+(p_P)} V_{\mu} \ket{{\bar B_s^0}(p_{B_s})}\;&=&\;F_1 (q^2) \;
(p_{B_s}+p_P) +
\;{m_{B_s}^2 - m_P^2 \over q^2} q_{\mu} \; \lbrack F_0 (q^2) - F_1 (q^2)
\rbrack\label{eq:3.1}\\
\bra{V^+(p_V)} J_{\mu} \ket{{\bar B_s^0}(p_{B_s})}\;&=&\;{2 V(q^2)\over m_{B_s}
+ m_V} \epsilon_{\mu \alpha \rho \sigma} \epsilon^{\ast \alpha} p_{B_s}^{\rho}
p_V^{\sigma} -\nn \\
&-&\;i \lbrack (m_{B_s} + m_V) A_1 (q^2) \epsilon_{\mu}^{\ast} - {A_2 (q^2)
\over m_{B_s} + m_V} (\epsilon^{\ast}\cdot p_{B_s}) (p_{B_s} + p_V)
_\mu -\nn \\
&-&\;(\epsilon^{\ast}\cdot p_{B_s}) {2 m_V \over q^2} q_\mu (A_3 (q^2) - A_0 (
q^2))\rbrack  \label{eq:vector}
\eea
\noindent
where $q^2 = (p_{B_s} - p_{P,V})^2$ and $J_\mu = {\bar q} \gamma_\mu (1 -
\gamma_5) b$ \hskip 5pt ($q=c,u$); $\epsilon$ is the $V^+$ meson polarization
 vector.
The conditions
\bea
F_1 (0)\;&=&\;F_0 (0)\nn \\
A_3 (0)\;&=&\;A_0 (0)\label{eq:relations}
\eea
\noindent
must be implemented in
eqs.(\ref{eq:3.1},\ref{eq:vector}) in order to avoid unphysical poles at
$q^2=0$;  $A_3$ can be expressed in terms of $A_1$ and $A_2$:
\be
A_3 (q^2)\;=\;{m_{B_s}+m_V \over 2 m_V} A_1 (q^2) - {m_{B_s}-m_V \over 2 m_V}
A_2 (q^2) \hskip 5pt . \label{eq:a3}\ee
\noindent
In the limit of massless charged leptons the relevant form factors are $F_1$,
$V$, $A_1$ and $A_2$.
Their calculation by QCD sum rules \cite{ovchinnikov} can be done by
considering the three-point correlators
\be
\Pi_{\mu} (p_{B_s}, p_P, q)\; = \;(i)^2 \int dx\;dy\;e^{i(p_P x - p_{B_s} y)}
\bra{0} T(J^P (x) V_\mu (0) J^{B_s \dag} (y))\ket{0} \label{eq:corrscal}
\ee
\noindent and
\be
\Pi_{\mu\nu}^{V,A}\;=\;(i)^2 \int dx\;dy\;e^{i(p_V x - p_{B_s} y)}
\bra{0} T(J_\nu^V (x) J_\mu^{V,A} (0) J^{B_s \dag} (y))\ket{0}
\label{eq:corrvect}
\ee
\noindent
where $J^{B_s}(y)={\bar s}(y) i \gamma_5 b(y)$, $J^P (x)={\bar s}(x) i
\gamma_5 q(x)$, $J_\nu^V (x)={\bar s}(x) \gamma_\nu q(x)$.
For $q=c$ the last two currents interpolate the $D_s^+$ and $D_s^{\ast +}$
meson respectively, whereas
for $q=u$, $J^V_\nu$ interpolates $K^{\ast +}$.

The correlators in (\ref{eq:corrscal},\ref{eq:corrvect}) can be
decomposed in Lorentz invariant structures:
\bea
\Pi_\mu (p_{B_s},p_P,q)\;&=&\;(p_{B_s}+p_P)_\mu \;\Pi\; + \;
(p_{B_s}-p_P)_\mu \; \Pi^{\prime} \label{eq:lorentz1}\\
\Pi_{\mu\nu}^V (p_{B_s},p_V,q)\;&=&\;\epsilon_{\mu\nu\rho\sigma}
p_V^{\rho} p_{B_s}^\sigma \; \Pi_V \label{eq:lorentz2}\\
\Pi_{\mu\nu}^A (p_{B_s},p_V,q)\;&=&\;i\lbrack g_{\mu\nu} \; \Pi_1 -
(p_{B_s}+p_V)_\mu {p_{B_s}}_\nu \; \Pi_2
- (p_{B_s}-p_V)_\mu {p_{B_s}}_\nu \; \Pi_3 -\nn \\
&-&\;{p_V}_\mu (p_{B_s}+p_V)_\nu \;\Pi_4 - {p_V}_\mu (p_{B_s}-p_V)_\nu \;\Pi_5]
\label{eq:lorentz3}
\eea
The saturation of the $p_{B_s}$ and $p_{P,V}$ channels by hadronic states
provides the hadronic side of the sum rules. For the invariant structures $\Pi$
, $\Pi_V$, $\Pi_1$ and $\Pi_2$ the following expressions can be written,
keeping the contribution of the lowest lying resonances only:
\bea
\Pi^{H}\;&=&\;({f_{B_s} m_{B_s}^2 \over m_b + m_s}) \; ({f_P m_P^2 \over
m_q + m_s}) \; F_1 (q^2) \;
 {1\over p_{B_s}^2 - m_{B_s}^2 + i\epsilon} \; {1\over
p_P^2 - m_P^2 + i\epsilon}\label{eq:hadron1}\\
\Pi_V^{H}\;&=&\;({f_{B_s} m_{B_s}^2 \over m_b + m_s}) \;
{m_V^2\over g_V} \; {2 V(
q^2) \over m_{B_s} + m_V} \; {1\over p_{B_s}^2 - m_{B_s}^2 + i\epsilon} \;
{1\over p_V^2 - m_V^2 + i\epsilon}\label{eq:hadron2}\\
\Pi_1^{H}\;&=&\;({f_{B_s} m_{B_s}^2 \over m_b + m_s}) \; {m_V^2\over g_V} \;
(m_{B_s} + m_V) A_1 (q^2) \;
{1\over p_{B_s}^2 - m_{B_s}^2 + i\epsilon} \;{1\over
p_V^2 - m_V^2 + i\epsilon}\label{eq:hadron3}\\
\Pi_1^{H}\;&=&\;({f_{B_s} m_{B_s}^2 \over m_b + m_s}) \;
{m_V^2\over g_V} \; {A_2 (q^2) \over m_{B_s} + m_V} \;
{1\over p_{B_s}^2 - m_{B_s}^2 + i\epsilon}\; {1\over
p_V^2 - m_V^2 + i\epsilon},\label{eq:hadron4}
\eea
\noindent
where $<0|J^V_\mu|V(p_V,\epsilon)>= (m^2_V/g_V) \; \epsilon_\mu$.
On the other hand, the correlators can be computed, for $p_{B_s}^2, p_{P,V}^2
\to - \infty$, by an operator product expansion in QCD in terms
of a perturbative
contribution and non perturbative power corrections.
For example, the perturbative contribution to $\Pi$ in eq. (\ref{eq:lorentz1})
reads:
\be
\Pi^{P}(p_{B_s}^2,p_P^2,q^2)\;=\;{1\over \pi^2} \int ds\;ds^{\prime}
{\rho_P (s^{\prime},s,q^2) \over (s^{\prime}-p^2_{B_s})(s-p^2_P)}
\label{eq:pertdisp}
\ee
\noindent where:
\bea
\rho_P(s,s^{\prime},q^2) &=& {3\over 2\chi^{3\over 2}}\lbrace {\chi \over 2}
 (\Delta + \Delta') - \chi \; m_s \;( 2 m_s - m_b - m_u ) \\
 & - &
\lbrack 2(s\Delta^{\prime}+s^{\prime}\Delta)
-u (\Delta+\Delta^{\prime})\rbrack \\
 & \times &\lbrack m_s^2-{u\over 2}+m_b m_q-m_q m_s-m_b m_s \rbrack \rbrace
\label{eq:density}
\eea
\noindent
with $\Delta=s-m^2_q+m^2_s$, $\Delta^\prime=s'-m^2_b+m^2_s$,
$\chi=(s + s^\prime + q^2) - 4 s s^\prime$ and $u=s+s^\prime + q^2$;
the integral in $s,s^\prime$ is within the domain bordered by the curves:
\be
s(s')_{\pm} = {2s^{\prime}(m_q^2-m_s^2)-s^{\prime}(m_b^2-m_s^2-s^{\prime})
\pm s^{\prime} \sqrt{(m_b^2 - m_s^2 -
s^{\prime})^2 - 4s^{\prime} m_s^2}\over 2s^{\prime}+(m_b^2 -
m_s^2 - s^{\prime}) \pm \sqrt{(m_b^2 - m_s^2 - s^{\prime})
^2 - 4s^{\prime} m_s^2}}
\label{eq:domain}
\ee
The power corrections to  $\Pi$ \cite{foot4},
 given in terms of quark and gluon condensates, read:
\bea
\Pi^{NP}&=&-{<{\bar s}s> \over 2 r r^{\prime}} (m_b + m_q)
+\nn \\
&+& \lbrack -m_s^2 <{\bar s}s> + {1\over 2} <{\bar s}\sigma \; g \; G s>
\rbrack
\lbrack {m_b^2 (m_b + m_q) \over 2 r r^{\prime 3}} - {m_q \over 4 r r^{\prime
2}} + {m_q^2 (m_b + m_q)\over 2 r^3 r^{\prime}}-\nn \\
&-&\;{m_b\over 4 r^2 r^{\prime}} + {(m_b + m_q)(m_b^2+m_q^2+Q^2)\over 4 r^2
r^{\prime 2}} + {m_b\over 4 r^2 r^{\prime}} + {m_q\over 4 r r^{\prime
2}}\rbrack+\nn \\
&+&\; {<{\bar s} \sigma \; g \; G s>\over 24} \lbrack {4 (m_b + 2 m_q)\over r^2
r^{\prime}} + {4 (m_q + 2 m_b)\over r r^{\prime 2}} + {(m_b + m_q)\lbrack (m_b
- m_q)^2 + Q^2\rbrack \over r^2 r^{\prime 2}} \nn \\
\label{eq:condensates}
\eea
\noindent
where $r=p_P^2 - m_q^2$ and $r^{\prime}=p_{B_s}^2 - m_b^2$.
The perturbative spectral densities $\rho_V$, $\rho_1$ and $\rho_2$ and the
power corrections to $\Pi_V$, $\Pi_1$ and $\Pi_2$ can be found in the
 appendix.
\par
We improve the matching between the hadronic side and the
QCD side of the sum rule
\be \Pi^{H}=\Pi^{P} + \Pi^{NP}\label{sumrule}\ee
by performing a double Borel transform to the variables $M^2$
and $M^{\prime 2}$ (conjugated to $-p_{P,V}^2$ and $-p_{B_s}^2$).
This suppresses the higher order power corrections in
the QCD side of the sum rule by factorials,
and enhances the contribution of the lowest
lying resonances in the hadronic side. By requiring stability in the variables
 $M^2$,
$M^{\prime 2}$ and hierarchy in the power corrections and in the
resonance-continuum contributions, a prediction for the form factors at
$Q^2=0$ can be obtained.
The quark masses, the condensates and the
effective thresholds are the same as in the previous section
(for $B_s\to K^{\ast}$ we use $s_0=1.2\div 1.3 \; GeV^2$);
as for the leptonic constants of the vector mesons, we use
$g_{D^*_s}=8.3$ (from the relation
$f_{D_s}/f_D=[m_{D^*_s}/g_{D^*_s}]/[m_{D^*}/g_{D^*}]$, with $g_{D^*}=7.8$)
and $g_{K^*}=4.3$.
\par
The results for the form factors of the transitions $B_s \to D_s, D_s^*$
at $Q^2=0$ are collected in Table I
($F_0(0)$,  $A_0(0)$ and $A_3(0)$
are obtained by eqs.(\ref{eq:relations},\ref{eq:a3})).
One can see that these values qualitatively agree with
the predictions of the BSW model \cite{WSB}.
As for the $Q^2$ dependence, it can be obtained in principle by QCD sum rules.
However, to avoid the relevant numerical uncertainties we prefer to
assume a polar dependence dominated by the nearest
resonance.  These resonances
are $\bar b c$  mesons  whose mass, for the lowest lying states, have been
estimated in \cite{colangelo3};  the $0^+$ and $1^+$ states are  $500 \;
MeV$ above $0^-$ and $1^-$, as suggested by the splitting between $S$ and $P$
states in the $D$ channel. In any case,
the results for the semileptonic widths,
as well as for the non leptonic widths calculated in the following section, are
quite insensitive to the exact position of the poles.
As for the Cabibbo suppressed transition
$B_s \to K^*$, the results for the form factors at $Q^2=0$ are:
$V(0)=0.12\pm 0.02$, $A_1(0)=0.3\pm 0.1$ and $A_2(0)\simeq 0$; however, a test
of the predictions based on these form factors is difficult.
\par
Using the form factors in Table I we predict, for
$V_{cb}=0.045$:
\bea
\Gamma(\overline {B_s^0} \to D^{+}_s \ell^- \bar\nu)&=&(1.35\pm 0.21)\cdot
10^{-14}\;GeV\;,\label{gamma1}\\
\Gamma(\overline  {B_s^0} \to D^{*+}_s \ell^- \bar\nu)&=&(2.5\pm 0.1)\cdot
10^{-14}\;GeV\;.\label{gamma2}
\eea

An estimate of the $SU(3)_F$ breaking effects can be
obtained by studying  the ratios
$F_1(B_s \to D_s) / F_1(B \to D)$, etc., with the result:
\bea
{F_1(B_s\to D_s)\over F_1(B\to D)}&=&1.12\pm0.04\label{eq:ffratio1}\\
{V(B_s\to D_s^{\ast})\over V(B\to D^{\ast})}&=&1.3\pm0.1
\label{eq:ffratio2}\\
{A_1(B_s\to D_s^{\ast})\over A_1(B\to D^{\ast})}&=&0.9\pm0.1
\label{ffratio3}\\
{A_2(B_s\to D_s^{\ast})\over A_2(B\to D^{\ast})}&=&1.3\pm0.1
\hskip 5pt . \label{ffratio4}
\eea

\section{ Two-body non leptonic $B_s$ decays}
We consider the two-body non leptonic ${\bar B_s^0}$ decays induced by
the effective weak hamiltonian
\be H_W={G_F\over \sqrt{2}} V_{cb} V_{q_2 q_1}^{\ast} \lbrack c_1 \;
({\bar c}b)_L
({\bar q_1}q_2)_L + c_2 \; ({\bar c}q_2)_L ({\bar q_1}b)_L \rbrack \hskip 5pt .
\label{hw}\ee
The Wilson coefficients $c_1$ and $c_2$, evaluated at the $b$-quark mass
scale $m_b\simeq 5\;
GeV$, are given by \cite{petrarca}:
\be c_1 (m_b)=1.1\;,\;\;\;\;\;\;\;c_2 (m_b)=-0.24 \hskip 5pt.\label{eq:4.2}\ee
The usual way to evaluate the matrix element of the operator
(\ref{hw}) between the ${\bar B_s^0}$ state and, e.g., the $D_s^+
\pi^-$ state is to assume a factorization in the product of the
$\bra{D_s^+} ({\bar c}b)_L \ket{{\bar B_s^0}}$
matrix element and the $\bra{\pi^-} ({\bar d}u)
_L \ket{0}$ matrix element. One obtains
\be
\bra{D_s^+ \pi^-} H_W \ket{{\bar B_s^0}} = {G_F\over \sqrt{2}}
V_{cb} V_{q_2 q_1}^{\ast}  a_1
\bra{D_s^+ } ({\bar c}b)_L  \ket{{\bar B_s^0}}
\bra{\pi^- } ({\bar d}u)_L  \ket{0} \hskip 5pt ,  \label{eq:4.3}\ee
\noindent
with $a_1 = c_1 + c_2/N_c$ ($N_c$ is the number of colours).
In this way  the non leptonic amplitude is given in terms of the semileptonic
matrix element parameterized in eq.(\ref{eq:3.1}) and of the pion
leptonic constant $f_\pi = 132 \; MeV$.
 As for the coefficient $a_1$, the analysis of non leptonic
$B_{u,d}$ decays  shows that the rule of discarding $1/ N_c$ corrections
should be adopted \cite{BSW};  we follow this rule and use $a_1=c_1=1.1$.
The relevant
leptonic constants are the same as in section 2, or they are fixed from the
experimental data. The
resulting non leptonic widths for several two body
${\bar B_s^0}$ decays are collected in Table II;  the
branching ratios  in the same Table are obtained using $\tau_{B_s}=1.2 \; ps$
\cite{bijens}. \par
It is worth observing that the channels with largest branching ratio, e.g.
${\bar B_s^0}\to D_s^+ \pi^-$ or
${\bar B_s^0}\to D_s^+ D_s^-$, could be revealed in
the LEP experiments \cite{oliver}.

\section{ Conclusions}
We have studied several aspects of the $B_s$ meson phenomenology by
QCD sum rules. Our main result concerns the possibility of
obtaining the size of the $SU(3)_F$ breaking effects
in this channel; we have shown
that the method is sensitive to such effects and can predict them carefully.

The deviation of $f_{B_s}$ from  the leptonic constant of the $B_d$ meson is
around $10 \%$; such deviation is of the same order  as predicted
by the Heavy Quark Effective Chiral Theory \cite{grinstein} but its origin
is different since in the QCD sum rules approach
it must be ascribed to the finite strange quark
mass and to the value of the strange quark condensate,  whereas in
\cite{grinstein} the deviation is connected to chiral loops.
$SU(3)_F$ breaking effects are at $10 - 20 \%$ level in the semileptonic form
factors; it should be interesting to compare this result with the prediction of
the Heavy Quark Effective Chiral Theory.

Finally, we have calculated the width of several non leptonic $B_s$
decays; some of them are in the LEP discovery potential.

\newpage
\renewcommand{\thesection}{\Alph{section}}
\setcounter{section}{1}
\setcounter{equation}{0}
\section*{{\bf Appendix}}
The perturbative spectral densities $\rho$ in eqs. (\ref{eq:pertdisp})
can be computed by applying the Cutkosky rule
\be {i\over k^2 - m^2}\;\to\;2 \pi \delta_+ (k^2 - m^2)\label{cutkosky}\ee
\noindent
to the triangle diagrams corresponding to the three point functions in eqs.
(\ref{eq:corrscal}) and (\ref{eq:corrvect}).
For the vector and axial current correlators these spectral densities are as
follows:
\bea
\rho_V (s,s^{\prime},q^2) &=& {3\over \chi^{3\over 2}}\lbrace (2s^{\prime}
\Delta - u\Delta^{\prime})(
m_s - m_q) + (2s\Delta^{\prime} - u\Delta)(m_s - m_b) + m_s\chi
\rbrace\label{eq:vector1}
\eea

\bea
\rho_1 (s,s^{\prime},q^2) &=& {3\over \chi^{1\over 2}} \lbrace (m_b - m_s)
\lbrack m_s^2 + {1\over \chi}(s^{\prime}\Delta^2+s{\Delta^{\prime}}^2-u\Delta
\Delta^{\prime})\rbrack-m_q (m_s^2-{\Delta^{\prime}\over 2})-\nn \\
&-&m_b (m_s^2-{\Delta\over 2})+m_s \lbrack m_s^2-{1\over 2}(\Delta+
\Delta^{\prime}-u)+m_b m_q \rbrack \rbrace\label{eq:axial}
\eea

\bea
\rho_2 (s,s^{\prime},q^2) &=& {3\over 2\chi^{3\over 2}} \lbrace m_b \lbrack
2s\Delta^{\prime}-u\Delta+4\Delta\Delta^{\prime}+2\Delta^2\rbrack+m_b m_s^2
(4s-2u)+\nn \\
&+&m_q (2s^{\prime}\Delta-u\Delta^{\prime})-m_s \lbrack 2(3s\Delta^{\prime}
+s^{\prime}\Delta)-u(3\Delta+\Delta^{\prime}
)+\nn \\
&+&\chi+4\Delta\Delta^{\prime}+2\Delta^2+m_s^2 (4s-2u)\rbrack +
{6\over \chi}(m_b-m_3)
\lbrack 4ss^{\prime}\Delta\Delta^{\prime}-\nn \\
&-&u(2s\Delta\Delta^{\prime}+s^{\prime}\Delta^2+s{\Delta^{\prime}}^2)+
2s(s^{\prime}\Delta^2+s{\Delta^{\prime}}^2)\rbrack\rbrace
\label{axial1}
\eea
\noindent
The non perturbative power corrections can be computed by applying
the fixed point technique \cite{novikov}.
The result is:

\bea
\Pi_V^{<{\bar s}s>} &=& -<{\bar s}s> \lbrace {1\over r r^{\prime}}
-{2m_b^2 m_s^2\over r {r^{\prime}}^3}-{2m_q^2 m_s^2\over r^3 r^{\prime}}
+{m_s^2 (m_b^2+m_q^2+Q^2)\over r^2{r^{\prime}}^2}\rbrace\label{eq:vcond3}
\eea

\bea
\Pi_V^{<{\bar s}\sigma G s>} &=& {1\over 6}  <{\bar s}\sigma \; g \; G s>
\lbrack {3 m_q^2\over r^3 r^{\prime}} + {3 m_b^2\over r {r^{\prime}}^3}
- {2\over r {r^{\prime}}^2}+\nn \\
&+& {1\over r^2 {r^{\prime}}^2}(2 m_q^2 + 2 m_b^2 - m_q m_b + 2 Q^2)\rbrack
\label{eq:vcond5}
\eea

\bea
\Pi_1^{<{\bar s}s>} &=& - <{\bar q_3}q_3> \lbrace {1\over r r^{\prime}}
 \lbrack
{1\over 2}((m_b+m_q)^2+Q^2)-{m_s^2\over 2}\rbrack +{1\over 2r}+\nn \\
&+&{1\over 2r^{\prime}}+{m_s^2\over 4}{1\over r^2 r^{\prime}}
((m_b+m_q)^2+Q^2)+{m_s^2\over
4}{1\over r {r^{\prime}}^2}((m_b+m_q)^2+Q^2)+\nn \\
&+&{m_q^2 m_s^2\over r^3 r^{\prime}}(m_b m_q+m_b^2+m_q^2+Q^2)+{m_b^2 m_s^2
\over r {r^{\prime}}^3}
(m_b m_q+m_b^2+m_q^2+Q^2)+\nn \\
&+&{m_s^2\over 4}{m_b^2+m_q^2+Q^2\over r^2 {r^{\prime}}^2}
((m_b+m_q)^2+Q^2)-{m_s^2
m_q^2\over 2r^3}-{m_s^2 m_b^2\over 2{r^{\prime}}^3}\rbrace
\label{eq:axcond3}
\eea

\bea
\Pi_1^{<{\bar s}\sigma G s>} &=& {1\over 12}<{\bar s}\sigma \; g \; G s>
\lbrace {3m_q^2\over r^3 r^{\prime}}(m_q^2+m_b^2+2m_b m_q+Q^2)+\nn \\
&+&{3m_b^2\over r {r^{\prime}}^3}(m_b^2+m_q^2+2m_b m_q+Q^2)+{1\over r^2
{r^{\prime}}^2}\lbrack 3m_b
m_q(m_b^2+m_q^2+Q^2)+\nn \\
&+&2((m_b^2+m_q^2+Q^2)^2-m_b m_q)\rbrack +{1\over r^2 r^{\prime}}
\lbrack 3m_q (m_b+m_q)
+2(m_b^2+Q^2)\rbrack\nn \\
&+&{1\over r {r^{\prime}}^2}\lbrack 3m_b 3m_q+m_b)+4(m_q^2+Q^2)\rbrack
-{2\over rr^{\prime}}+\nn \\
&+&{3m_q^2\over r^3}+{3m_b^2\over {r^{\prime}}^3}+{2\over{r^{\prime}}^2}\rbrace
\label{eq:axcond5}
\eea

\bea
\Pi_2^{<{\bar s}s>} = - {1\over 2}<{\bar s}s> \lbrace {1\over r
r^{\prime}}+{m_s^2 m_q^2\over r^3 r^{\prime}}+{m_s^2 m_b^2\over r
{r^{\prime}}^3}+{(m_b^2+m_q^2+Q^2)
m_s^2\over 2r^2 {r^{\prime}}^2}-{m_s^2\over r {r^{\prime}}^2}\rbrace
\label{eq:ax1cond3}
\eea

\bea
\Pi_2^{<{\bar s}\sigma G s>} &=& {1 \over 12} <{\bar s}\sigma \; g \; G s>
\lbrace {3m_q^2\over r^3 r^{\prime}}+{3m_b^2\over r {r^{\prime}}^3}
-{2\over r {r^{\prime}}^2}+{1\over r^2
{r^{\prime}}^2}(2m_q^2+\nn \\
&+&2m_b^2+2Q^2-m_b m_q)\rbrace
\label{eq:ax1cond5}
\eea

\newpage
\vskip 1cm
\begin{center} {\bf Table captions} \end{center}
\vskip 1cm
\noindent {\bf Table I.} Values at $q^2=0$
of the form factors appearing in the matrix
elements of the decays
${\bar B_s^0} \to D_s^+ (D_s^{*+}) \ell^- {\bar \nu_\ell}$.
The quantum numbers and the mass of the poles which determine the $q^2$
dependence of the form factors is also shown.

\vskip 1cm
\noindent {\bf Table II.} Two-body non leptonic
${\bar B_s^0}$ decay widths. The branching ratios are obtained for
$\tau_{B_s}=1.2 \; ps$.

\newpage
\vskip 1cm
\begin{center} {\bf Figure captions} \end{center}
\vskip 1cm
\noindent {\bf Fig. 1.} Stability analysis for the mass and the leptonic
constant of the $B_s$ meson. The solid line corresponds to $s_0=33$ $GeV^2$,
the dashed line to $s_0=34$ $GeV^2$, the dotted line to $s_0=35$ $GeV^2$ and
the dashed-dotted line to $s_0=36$ $GeV^2$. $M$, $f_{B_s}$ and $m_{B_s}$ are in
$GeV$.
\vskip 1cm

\noindent {\bf Fig. 2.} Stability analysis for the ratios ${m_{B_s}\over
m_{B_d}}$ and ${f_{B_s}\over f_{B_d}}$. The symbols are the same as in fig. 1.
\newpage
\begin{table}
\begin{center}
\begin{tabular}{c c c c}
{\bf Table I}\\ \\
\hline  \hline \\
$B_s \to D_s, D^*_s$
 & value at $q^2=0$ & $J^P$ of the pole & pole mass (GeV) \\ \\
\hline \hline \\
$F_1$ & $0.7\pm 0.1$ & $1^{-}$ & 6.3 \\ \\
$F_0$ & $0.7\pm 0.1$ & $0^{+}$ & 6.8 \\ \\
$V$ & $0.63\pm 0.05$ & $1^{-}$ & 6.3 \\ \\
$A_0$ & $0.52\pm 0.06$ & $0^{-}$ & 6.3 \\ \\
$A_1$ & $0.62\pm 0.01$ & $1^{+}$ & 6.8 \\ \\
$A_2$ & $0.75\pm 0.07$ & $1^{+}$ & 6.8 \\ \\
$A_3$ & $0.52\pm 0.06$ & $1^{+}$ & 6.8 \\ \\ \hline \hline \\
\end{tabular}
\end{center}
\end{table}
\newpage

\begin{table}
\begin{center}
\begin{tabular}{l c c}
{\bf Table II}\\ \\
\hline  \hline \\
Decay mode & Width$\times ({V_{cb}\over 0.045})^2$ (GeV) &
Branching ratio \\ \\ \hline \hline \\
${\bar B_s^0}\to D_s^{\ast +} D_s^{\ast -}$
& $9\times 10^{-15}$ & $1.6\times
10^{-2}$ \\ \\
${\bar B_s^0}\to D_s^{\ast +} \rho^{-}$ & $7\times 10^{-15}$ & $1.3\times
10^{-2}$ \\ \\
${\bar B_s^0}\to D_s^{\ast +} D^{\ast -}$ & $5\times 10^{-16}$ & $8\times
10^{-4}$ \\ \\
${\bar B_s^0}\to D_s^{\ast +} K^{\ast -}$& $4\times 10^{-16}$ & $6\times
10^{-4}$ \\ \\
${\bar B_s^0}\to D_s^{+} \rho^{-}$ & $7\times 10^{-15}$ & $1.3\times
10^{-2}$ \\ \\
${\bar B_s^0}\to D_s^{+} a_1$ & $6\times 10^{-15}$ & $1.1\times
10^{-2}$ \\ \\
${\bar B_s^0}\to D_s^{+} D_s^{-}$ & $6\times 10^{-15}$ & $1\times
10^{-2}$ \\ \\
${\bar B_s^0}\to D_s^{+} D_s^{\ast -}$ & $4\times 10^{-15}$ & $8\times
10^{-3}$ \\ \\
${\bar B_s^0}\to D_s^{+} \pi^{-}$ & $3\times 10^{-15}$ & $5\times
10^{-3}$ \\ \\
${\bar B_s^0}\to D_s^{\ast +} \pi^{-}$ & $1\times 10^{-15}$ & $2\times
10^{-3}$ \\ \\
${\bar B_s^0}\to D_s^{\ast +} D_s^{-}$ & $2\times 10^{-15}$ & $4\times
10^{-3}$ \\ \\
${\bar B_s^0}\to D_s^{\ast +} K^{-}$ & $1\times 10^{-16}$ & $2\times
10^{-4}$ \\ \\
${\bar B_s^0}\to D_s^{\ast +} D^{-}$ & $1\times 10^{-16}$ & $2\times
10^{-4}$ \\ \\
${\bar B_s^0}\to D_s^{+} D^{-}$ & $3\times 10^{-16}$ & $5\times
10^{-4}$ \\ \\
${\bar B_s^0}\to D_s^{+} K^{-}$ & $2\times 10^{-16}$ & $4\times
10^{-4}$ \\ \\
${\bar B_s^0}\to D_s^{+} D^{\ast -}$ & $2\times 10^{-16}$ & $4\times
10^{-4}$ \\ \\
${\bar B_s^0}\to D_s^{+} K^{\ast -}$ & $4\times 10^{-16}$ & $6\times
10^{-4}$ \\ \\ \hline \hline
\end{tabular}
\end{center}
\end{table}

\begin{thebibliography}{99}
\bibitem{foot1} In the present paper we shall consider both the charge
conjugate $B_s$ states.
\bibitem{foot4}
We only consider the corrections that survive after the double Borel
transform.
\end{thebibliography}
\end{document}